\documentclass[pre,twocolumn,superscriptaddress]{revtex4}
\usepackage{graphicx}
\usepackage{graphics}
\usepackage{latexsym}
\usepackage{amsmath, amsthm, amssymb, wasysym}
\usepackage{hyperref,xcolor}
\usepackage{epsfig}
\usepackage[mathscr]{euscript}

\begin{document}

\title{Capture of a diffusing lamb by a diffusing lion when both return
home}
\author{R. K. Singh}
\email{rksinghmp@gmail.com}
\affiliation{Department of Physics, Bar-Ilan University, Ramat-Gan 5290002, Israel}

\author{Sadhana Singh}
\email{sdhnsingh080@gmail.com}
\affiliation{The Avram and Stella Goldstein-Goren Department of Biotechnology Engineering, 
Ben-Gurion University of the Negev, Be'er Sheva 84105, Israel}

\begin{abstract}
A diffusing lion pursues a diffusing lamb when both of them are allowed to get
back to their homes intermittently. Identifying the system with a pair of vicious
random walkers, we study their dynamics under Poissonian and sharp resetting.
In absence of any resets, the location of intersection of the two walkers
follows a Cauchy distribution. In presence of resetting, the distribution of the
location of annihilation is composed of two parts: one in which the trajectories
cross without being reset (center) and the other where trajectories are reset
at least once before they cross each other (tails). We find that the tail part
decays exponentially for both the resetting protocols. The
central part of the distribution, on the other hand, depends on the nature
of the restart protocol, with Cauchy for Poisson resetting and Gaussian
for sharp resetting. We find good agreement of the analytical results with numerical
calculations.
\end{abstract}

\maketitle

\newcommand{\fr}{\frac}
\newcommand{\tl}{\tilde}
\newcommand{\avgr}{\langle \mathscr{T}_R \rangle}
\newcommand{\avgt}{\langle \mathscr{T}_T \rangle}

\section{Introduction}
Search is a fundamental endeavor to survival ranging from human search \cite{bell2012searching}
to rescue operations \cite{shlesinger2009random} to animal foraging \cite{bartumeus2009optimal,
benichou2006two}
to protein binding on DNA \cite{coppey2004kinetics}, transcription factors searching for
a specific DNA \cite{von2007simple, gorman2008visualizing}, to mention a few. A useful search
strategy involves intermittent phases of slow motion aiding the searcher in target
detection and fast motion allowing the searcher to cover maximal ground in minimal
time \cite{benichou2011intermittent}. Restarting a search process at intermittent intervals,
aka stochastic resetting, has been extensively shown to expedite search
\cite{evans2020stochastic}.
Stochastic resetting has been a very active topic of research within the realm of
nonequilibrium statistical physics over the past decade \cite{evans2020stochastic}. The basic
essence of stochastic resetting is that in any kind of search process, the search
is rarely successful in the first attempt. Following which the search is restarted again
and again until the process culminates with success. This property is common to
a wide variety of search processes. Now
it is almost always true that if sufficient amount of time is devoted,
then any search shall meet success. The question of value is, however,
whether an intermittent restart of the search process tends to reduce the time of
completion? Answer to this question is affirmative. At least in the case of 
stochastic algorithms it has been shown that a simple restart
might expedite completion \cite{luby1993optimal, tong2008random, avrachenkov2013markov}.
Not only in endeavors of human interest, nature also employs restarts in many processes, for example 
enzymatic reactions following the Michaelis-Menten reaction scheme \cite{michaelis1913kinetik}.

The idea of stochastic resetting to Brownian search problem was first applied in the seminal
work of Evans and Majumdar \cite{evans2011diffusion}. They showed that restarting
a Brownian particle to its initial location at a constant rate renders the mean first
passage time (MFPT) finite. The process of restarting a stochastic process at a
constant rate is termed as Poissonian resetting. In addition to Brownian motion, Poissonian
resetting has been applied to run and tumble particles \cite{evans2018run}, fluctuating interfaces
\cite{gupta2014fluctuating}, dynamical phase transitions \cite{majumdar2015dynamical,
singh2022general}, resetting transitions \cite{ahmad2019first, ahmad2022first},
telegraphic processes \cite{masoliver2019telegraphic}, comb-like
structures \cite{domazetoski2020stochastic, singh2021backbone}, multiple
Brownian searchers \cite{evans2014diffusion}, etc.
However, Poissonian resetting is not exclusive and other protocols like power-law
distributed resetting times \cite{nagar2016diffusion}, resetting rates depending
on space \cite{evans2011JPA} and time \cite{pal2016diffusion} have also been extensively
studied. This raises an interesting question: given the wide class of resetting
protocols, does there exist a reset mechanism under which MFPT is minimal?
This question is difficult to answer in its full generality. However, when resetting
is renewal, then sharp resetting in which the time interval between two resets is
fixed serves as the best strategy \cite{pal2017first,chechkin2018random}.
In other words, ``if there exists a stochastic resetting protocol
that improves search process, then there exists a deterministic restart protocol that
performs as good or better'' \cite{eliazar2020mean}.

Poissonian and sharp restarts lie at the two extremes of renewal resetting,
former being memoryless and the latter retaining its entire memory. Both these
protocols were compared against each other for a system of Brownian particles
searching for a target in Ref.~\cite{bhat2016stochastic} and it was shown that
sharp resetting typically leads to a lower search cost than that in Poissonian
resetting. This study was taken further for a system of Brownian
particles where interactions are relevant, for example, in population genetics
\cite{da2021diffusion}. Inclusion of interactions further allows to consider
more nontrivial forms of resetting mechanisms such as those which are driven by the
interactions between the constituent particles \cite{falcao2017interacting} or
space dependent resetting in interacting Brownian particles \cite{bertin2022stochastic}.
One of the most important examples which involves interacting random walks is the
well known prey predator model which culminates when the prey is captured by
the predator \cite{krapivsky1996kinetics}. An exactly solvable prey predator
model with resetting was recently considered by Evans and co-workers
\cite{evans2022exactly} where the prey on its encounter with a predator can either
perish or be reset to its initial location.

In the present work we consider the prey predator model within the realm of vicious
random walks which annihilate each other the moment their trajectories cross
\cite{bray2013persistence}. The concept was first introduced by Fisher in the context of
interfacial wetting in 1+1 dimensions \cite{fisher1984walks,huse1984commensurate} and
has since been applied to Coulomb gas \cite{forrester1989vicious} and random matrices
\cite{baik2000random}. The survival probability for vicious
random walkers in one dimension exhibits power law decaying tails \cite{bray2004vicious},
and any two such walkers shall certainly meet each other as a random walk in one dimension
is recurrent \cite{klafter2011first,weiss1983random}. The problem of reunion of
two vicious random walks corresponds to the following chemical reaction: $A + A
\rightarrow \phi$ \cite{cardy1996theory} and is one of the most classic problems in nonequilibrium
statistical physics \cite{fisher1988reunions, schehr2008exact, kundu2014maximal}.
The annihilating nature of the vicious walkers makes them suitable for studying
directed polymer brushes wherein the viciousness captures the role of the non-intersecting
property of polymers \cite{essam1995vicious}. Vicious random walks have also been
applied to breathing DNA with the collapse of the bubbles viewed as an annihilation
of two vicious walkers moving in opposite potentials \cite{pedersen2009bubble}.
Furthermore, the distribution of the location of coalescence makes it relevant
to study the location where the trajectories of two vicious walkers cross. In the context
of the capture problem where a hungry lion pursues a lamb \cite{redner1999capture},
the location of intersection tells us how far the hunt is made from the home.
The scenario also makes the concept of resetting very natural \cite{reuveni2016optimal}.
This is because either the lamb shall every now and then return to
its home, or the lion to its den, or both. The reason that such a thing might happen 
as the lion pursues the lamb but could not catch it and gets tired eventually
getting back in its cave. On the other hand, the freely roaming lamb might spot the lion
and run away from it. This makes the study of vicious random walks under resetting very
natural. In other words, if we have two vicious Brownian particles we want to know how
long do they survive without crossing each other's paths? And if their
trajectories cross, what is the nature of the distribution of such a point? Do the answers
to these questions depend on the resetting protocol employed? We address these questions
in the following sections by studying the system of two vicious Brownian particles under
resetting. The particles are reset identically to their respective initial positions either
at constant rates (Poissonian resetting) or after fixed time intervals (sharp resetting).

\section{Two vicious random walkers}
Consider two Brownian particles:
\begin{subequations}
\label{dyn}
\begin{align}
\dot{x}_1 &= \eta_1(t),\\
\dot{x}_2 &= \eta_2(t)
\end{align}
\end{subequations}
where $\eta_1(t), \eta_2(t)$ are independent Gaussian random deviates with mean zero and
delta correlated variance, that is, $\langle \eta_1(t) \eta_1(t') \rangle = 2
D_1\delta(t-t')$ and $\langle \eta_2(t) \eta_2(t') \rangle = 2
D_2\delta(t-t')$. At $t = 0$ the two walkers are at $x_1 = 0$ and $x_2 = L$.
The two walkers annihilate each other as soon as their paths cross, that is,
$x_1(t) = x_2(t)$. The problem is readily transformed to the motion of the center of
mass $x_c = \fr{x_1+x_2}{2}$ and relative separation of the two particles
$x_r = x_1 - x_2$. In terms of the new coordinates, the center of mass moves as
a free Brownian particle as
\begin{align}
\dot{x}_c(t) = \eta_c(t)
\end{align}
where $\langle \eta_c(t) \rangle = 0$ and $\langle \eta_c(t) \eta_c(t') \rangle = 2
D_c\delta(t-t')$ with $D_c = \fr{D_1+D_2}{4}$. On the other hand, the relative coordinate
$x_r$ moves like a Brownian particle on line
\begin{align}
\dot{x}_r(t) = \eta_r(t)
\end{align}
where $\langle \eta_r(t) \rangle = 0$ and $\langle \eta_r(t) \eta_r(t') \rangle = 2
D_r\delta(t-t')$ with $D_r = D_1+D_2$. Before the trajectories of the two particles
cross, the center of mass exhibits a Brownian motion centered at $x_c = L/2$ with a
diffusion coefficient $D_c$ and the relative coordinate is a Brownian particle starting
at $x_r = -L$ with an absorbing wall at $x_r = 0$. The first passage time distribution
(FPTD) of the relative coordinate to the absorbing wall at $x_r = 0$ is
$F(t) = \fr{L}{\sqrt{4\pi D_r t^3}}\exp\Big(
-\fr{L^2}{4D_r t}\Big)$ and the probability density function (PDF) of the center of
mass motion is $p(x_c,t) = \fr{1}{\sqrt{4\pi D_c t}}\exp\Big[-\fr{(x_c-L/2)^2}
{4D_c t}\Big]$ \cite{gardiner1985handbook,redner2001guide}. From the FPTD
it is evident that the mean time to the annihilation of the two vicious
walkers $\langle t \rangle = \int^\infty_0 dt ~t ~F(t)$ is infinite. The recurrence
of a Brownian motion in one dimension, however, implies that the two walkers will
eventually collide, and the PDF of the location of intersection is
\begin{align}
\label{cauchy}
h(x_c) &= \int^\infty_0 dt ~F(t) p(x_c,t)\nonumber\\
&= \fr{1}{\pi}\fr{L/\sqrt{D_r D_c}}{\fr{L^2}{D_r} + \fr{(x_c-L/2)^2}{D_c}},
\end{align}
which is a Cauchy distribution centered at $x_c = L/2$. Similar to the MFPT,
there is no well defined mean location
of the intersection of the two vicious walkers. This is because even though the two walkers
shall certainly meet, they may take really long time to do so by venturing out in opposite
directions resulting in the divergence of MFPT and a well defined mean location of annihilation.
In other words, the hungry lion may keep pursuing the lamb forever and might eventually die
of hunger. And this is where resetting comes in to prevent the hungry lion from dying.

\section{Resetting to initial configuration}
With the vector $(x_1,x_2)$ defining the system, define a resetting protocol: 
after an interval of reset time $\tau$ the system is reverted back to 
its initial configuration. The time $\tau$
is either an exponentially distributed random variable (Poissonian resetting) or a
fixed quantity (sharp resetting). For simplicity let us assume that
the two walkers are reset via identical resetting protocols at exact 
same time. The reason for this choice is the following: let us assume that the two
walkers are reset at different times
$\tau_1$ and $\tau_2$, then a scenario is possible in
which $\tau_1 < \tau_2$ and $x_1(\tau_1) < x_2(\tau_2) < 0$ just before reset has
taken place. The moment after the reset we have $x_1(\tau_1) > x_2(\tau_2)$ which
apparently means that the two trajectories have crossed paths. Such crossing of two walkers 
is, however, erroneous as at the time of reset the first particle is removed from
its current location and put back to its initial location instantaneously. This
makes intersection point an ill-defined quantity simply for the reason that
actual trajectories did not cross. We avoid such a pathological situation
by requiring that the two particles be reset at exact same time. Furthermore,
restarting the two walkers identically retains the advantage that the two particle system
is still described by the motion of the center of mass and motion about the center of
mass. Next we consider the two resetting protocols one by one.

\subsection{Poissonian resetting}
Let the two walkers be reset to their respective initial locations at a rate $R$. Then
the FPTD of the relative coordinate under Poissonian resetting is \cite{reuveni2016optimal}
\begin{align}
\label{F_ex}
\tl{F}_R(s) = \fr{\tl{F}(s+R)}{\fr{s}{s+R} + \fr{R}{s+R}\tl{F}(s+R)},
\end{align}
where $\tl{F}(s) = \int^\infty_0 dt~e^{-st} F(t) = \exp(-\sqrt{sL^2/D_r})$ is
the Laplace transform of the FPTD without resetting \cite{oberhettinger2012tables}.
From this follows the MFPT under Poissonian resetting as:
$\avgr = \fr{e^{\sqrt{RL^2/D_r}}-1}{R}$. This result has been previously derived in
\cite{evans2011diffusion} via the backward Fokker-Planck equation and many works following
it. Here we state the result as a reminder that MFPT under Poissonian resetting
is finite.

\begin{figure}
\includegraphics[width=0.5\textwidth]{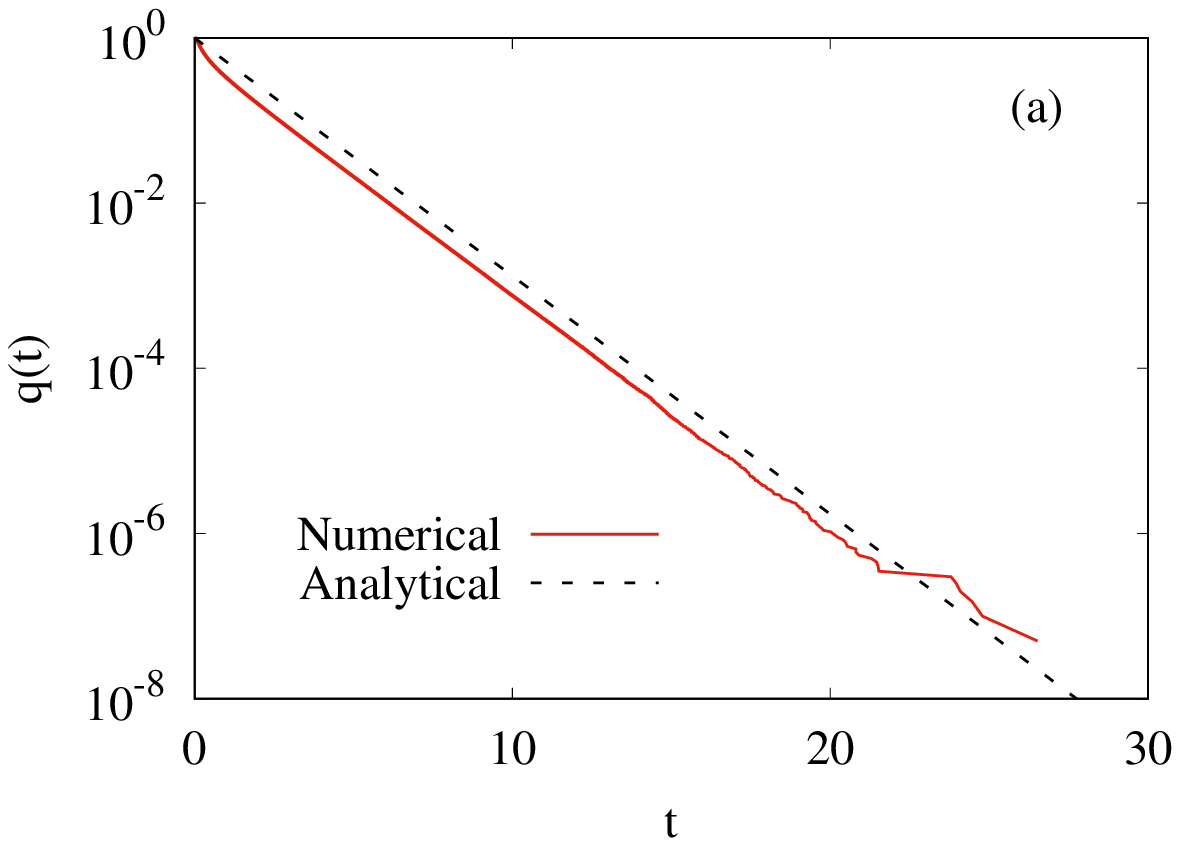}\\
\includegraphics[width=0.5\textwidth]{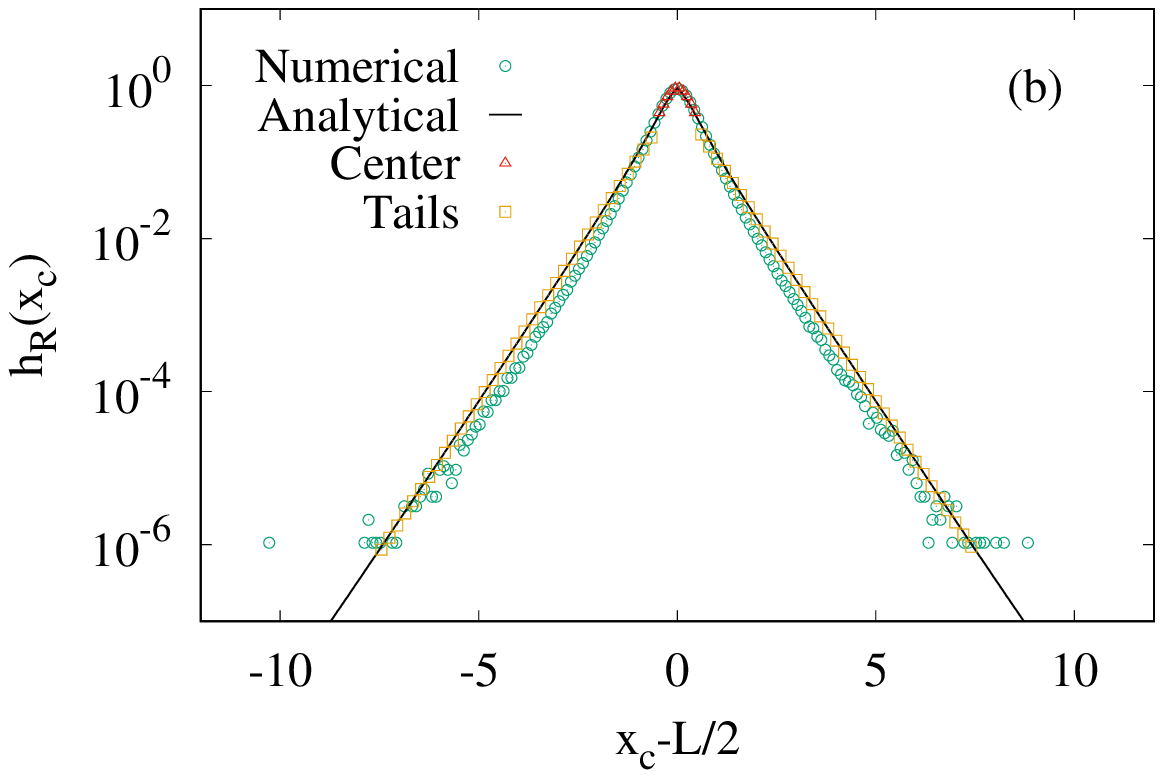}
\caption{(a) Survival probability $q(t)$ for the two vicious walkers
following Eq.~(\ref{dyn}) to not cross paths upto time $t$ reset to their initial
positions at a constant rate $R$. Red solid line represents
numerically estimated $q(t)$ while the black dashed line represents the
analytical form: $q(t) \sim \exp(-|s_{0,R}|t)$. (b) Numerically estimated PDF of
the location of intersection $h_R(x_c)$ (green circles) is compared against the
approximate form in Eq.~(\ref{app_hR}) (black solid line). Yellow squares denote the contribution
to $h^t_R$ for $|x_c-L/2|>L/2$ and red triangles $h^c_R$ for $|x_c-L/2|<L/2$.
We have used a factor $b$ such that $h_R \sim b h^c_R$ to demonstrate that
$h^c_R$ indeed captures the center of $h_R$ (upto a scale). Parameter values are:
$D_1, ~D_2 = 1,~L = 1,~R = 1$ and $b = 2$.}
\label{rate}
\end{figure}

In order to study the effect of resetting on the PDF of intersection 
of the two trajectories, we need the FPTD $F_R(t)$ given in terms of
the Bromwich integral \cite{arfken1999mathematical}
\begin{align}
F_R(t) &= \fr{1}{2\pi i}\int^{\gamma+i\infty}_{\gamma-i\infty}ds~
\fr{(s+R)e^{-\alpha L}}{s+Re^{-\alpha L}}~e^{st},\nonumber\\
&\stackrel{\text{large}~t}{\approx} \fr{1}{2\pi i}\int^{\gamma+i\infty}_{\gamma-i\infty}ds~
\fr{Re^{-z}}{s+Re^{-\alpha L}}~e^{st}
\end{align}
where $\alpha = \sqrt{\fr{s+R}{D_r}}$ and $z = \sqrt{R/D_r}L$. This integral is 
easily evaluated from the residue of $\tl{F}_R(s)$ at the pole closest to
$s = 0$. The pole of $\tl{F}_R(s)$ is given by the solution of $s+Re^{-\alpha L} = 0$
which in terms of $s = R(u-1)$ reads \cite{evans2011diffusion}
\begin{align}
u = 1 - e^{-\sqrt{u}z},
\end{align}
and has a unique nonzero solution $u_0 \in (0,1)$. Thus, the FPTD at large times is
\cite{evans2011diffusion}
\begin{align}
F_R(t) &\stackrel{\text{large}~t}{\approx} \lim_{s \to s_{0,R}} e^{st}(s-s_0)\tl{F}_R(s),\nonumber\\
&= \fr{2R\sqrt{u_0}e^{-z}}{2\sqrt{u_0} - z(1-u_0)}e^{s_{0,R} t},
\end{align}
where $s_{0,R} = R(u_0 - 1) < 0$ implying that at large times the FPTD under resetting
possesses exponentially decaying tails. 
This result is verified numerically and a good agreement is found for the 
characteristic decay exponent $s_{0,R}$ as shown in Fig.~\ref{rate}(a). 

Next, we estimate the PDF of the center of mass motion under resetting, following 
the renewal equation \cite{evans2020stochastic}
\begin{align}
\label{renewal}
p_R(x_c,t) = e^{-Rt}p(x_c,t) + \int^t_0 d\tau ~Re^{-R\tau} p(x_c,\tau),
\end{align}
where the first term gives the contribution from the trajectories which have not been
reset at all, while the second term describes the effect of resetting. As a result,
the PDF of the intersection point of the two vicious walkers under
Poissonian resetting annihilating each other with FPTD $F_R(t)$ is
\begin{align}
\label{exp_Poisson}
h_R(x_c) &= \int^\infty_0 dt ~F_R(t)p(x_c,t)e^{-Rt}\nonumber\\
&+ \int^\infty_0 dt ~F_R(t)\int^t_0 d\tau ~R e^{-R\tau} p(x_c,\tau)\nonumber\\
&\equiv h^c_R(x_c) + h^t_R(x_c),
\end{align}
where $h^c_R$ denotes the single integral and $h^t_R$ denotes the double
integral. In what follows we shall see that $h^c_R$ having the contribution of 
intersection points without reset captures the central part
of the PDF $h_R$. While $h^t_R$ describes the tails consisting the intersection 
points with reset, hence the usage of the
superscripts $c$ and $t$ respectively.
Let us now proceed to evaluate the two integrals in (\ref{exp_Poisson}) one by one.

For the single integral in Eq.~(\ref{exp_Poisson}), $F_R(t) = F(t) = \fr{L}
{\sqrt{4\pi D_r t^3}}\exp\Big(-\fr{L^2}{4D_r t}\Big)$, since the
system behaves like the one without resetting. Furthermore, if we look at the
FPTD under Poissonian resetting, that is, $\tl{F}_R(s) = \fr{(s+R)e^{-\alpha L}}
{s+Re^{-\alpha L}}$, then at small times $\tl{F}_R(s) \stackrel{\text{large}~s} \approx
\exp\Big[-L\sqrt{\fr{s+R}{D_r}}\Big] \Rightarrow F_R(t) \stackrel{\text{small}~t}
{\approx} F(t)e^{-Rt}$, which is exactly the same quantity as appearing in
the first integral in (\ref{exp_Poisson}). 
In other words, the probability of crossing of two trajectories without being reset is $e^{-Rt}$, 
effectively modifying the FPTD entering in
the evaluation of $h^c_R(x_c)$. Hence, we have
\begin{align}
\label{first_int}
&h^c_R(x_c)\nonumber\\
&= \fr{1}{\pi}\sqrt{\fr{RL^2}{L^2 D_c+(x_c-L/2)^2D_r}}K_1\Big[
\sqrt{R\Big\{\fr{L^2}{D_r} + \fr{(x_c-L/2)^2}{D_c}\Big\}}\Big]
\end{align}
where $K_1$ is the modified Bessel function of the second kind and it
enters while evaluating the Laplace transform of $e^{-1/t}$
\cite{oberhettinger2012tables}. The double integral in (\ref{exp_Poisson})
is similarly evaluated
\begin{align}
\label{second_int}
&h^t_R(x_c)\approx \int^\infty_0 dt ~A(s_{0,R}) e^{s_{0,R} t}
\int^t_0 d\tau ~Re^{-R\tau}p(x_c,t),\nonumber\\
&= \fr{RA(s_{0,R})}{\sqrt{4\pi D_c}}\int^\infty_0 d\tau~\fr{1}{\sqrt{\tau}}
e^{-R\tau-a/\tau}\int^\infty_\tau dt~e^{-|s_{0,R}|t},\nonumber\\
&= \fr{RA(s_{0,R})/|s_{0,R}|}{\sqrt{4D_c(|s_{0,R}|+R)}}
\exp\Big(-\sqrt{\fr{|s_{0,R}|+R}{D_c}}\Big|x_c-\fr{L}{2}\Big|\Big),
\end{align}
where $a = \fr{(x_c-L/2)^2}{4D_c}$ and $A(s_{0,R}) = \fr{2R\sqrt{u_0}e^{-z}}
{2\sqrt{u_0} - z(1-u_0)}$. Combining the results in (\ref{first_int}) and
(\ref{second_int}) we have  
\begin{widetext}
\begin{align}
\label{app_hR}
h_R(x_c) \approx \fr{1}{\pi}\sqrt{\fr{RL^2}{L^2 D_c+(x_c-L/2)^2D_r}}K_1\Big[
\sqrt{R\Big\{\fr{L^2}{D_r} + \fr{(x_c-L/2)^2}{D_c}\Big\}}\Big]
+ \fr{RA(s_{0,R})/|s_{0,R}|}{\sqrt{4D_c(|s_{0,R}|+R)}}
\exp\Big[-\sqrt{\fr{|s_{0,R}|+R}{D_c}}\Big|x_c-\fr{L}{2}\Big|\Big].
\end{align}
\end{widetext}
From Eq.~(\ref{app_hR}), it is evident that $h^c_R$ describes the center and $h^t_R$ the tails
of the PDF $h_R$. This is due to the rapid decay of the modified Bessel function
as compared to exponential. Furthermore, for small arguments $K_1$ decays
algebraically, that is, $K_1(w) \stackrel{\text{small}~w}{\sim}1/w$
\cite{abramowitz1988handbook}, from where it follows that
$h^c_R(x_c)$ behaves like a Cauchy distribution. Thus, when the two walkers
are reset to their initial locations at a constant rate, the PDF $h_R$
exhibits a Cauchy distributed center and exponentially
decaying tails. In other words, Poissonian resetting of the two walkers reduces
the fat tails of the PDF to the center
and the far tails are modified to exponential. In summary,
\begin{align}
h_R(x_c)\approx\begin{cases}
\fr{1}{\pi}\fr{L/\sqrt{D_r D_c}}{\fr{L^2}{D_r} + \fr{(x_c-L/2)^2}{D_c}},~\text{center},\\
\fr{RA(s_{0,R})/|s_{0,R}|}{\sqrt{4D_c(|s_{0,R}|+R)}}
\exp\Big[-\sqrt{\fr{|s_{0,R}|+R}{D_c}}\Big|x_c-\fr{L}{2}\Big|\Big]~\text{tails}.
\end{cases}
\end{align}
We compare the analytically estimated PDF $h_R$ in Eq.~(\ref{app_hR}) (black solid
line) with numerical calculations (green circles) in Fig.~\ref{rate}(b) and find
that they are in close proximity.

The contribution to the PDF coming from 
$h^c_R$ in the range $|x_c-L/2|<L/2$ is scaled by a factor $b$ to match
the numerically estimated $h_R$. The reason for doing this is to show that $h^c_R$
indeed captures the shape of the center of the PDF.
The tail part $h^t_R$ in region $|x_c-L/2|>L/2$ matches well with numerically estimated
$h_R$ as following the similar decay rate as shown in Fig.~\ref{rate}(b).
Here, we refer to the region $|x_c-L/2|<L/2$
as the central part. The simple reason for the usage of this terminology is that
the center of mass is midway between the two vicious particles and an annihilation
taking place within this region would simply mean that the center of mass has not
ventured far from its mean position.

\subsection{Sharp resetting}
In sharp resetting, the two walkers are reset to their respective initial locations after
fixed intervals of time $T$. In order to estimate the FPTD under sharp resetting, we
use the results derived by Pal and Reuveni in Ref.~\cite{pal2017first}. They show
that if $\tau$ is the time of completion of a stochastic process without restart, and $\rho$
is the time interval of restart, then the FPTD under restart reads \cite{pal2017first}
\begin{align}
\label{F_res}
\tl{F}_{res}(s) = \fr{\text{Pr}(\tau<\rho)\tl{\tau}_\text{min}(s)}
{1-\text{Pr}(\rho\le\tau)\tl{\rho}_\text{min}(s)},
\end{align}
where $\rho_\text{min} = \{\rho | \rho = \text{min}(\rho,\tau)\}$ is the random restart
time given restart occurred before completion and $\tau_\text{min} = \{\tau | \tau =
\text{min}(\rho,\tau)\}$ is the random completion time without any restarts. For Poissonian
resetting when $\rho$ is an exponentially distributed random variable, that is,
$f_\rho(t) = Re^{-Rt}$,
Eq.~(\ref{F_res}) reduces to (\ref{F_ex}) (see SM in Ref.~\cite{pal2017first}).
For sharp resetting at fixed intervals of time $T$, the distribution of restart
times $\rho$ is $f_\rho(t) = \delta(t-T)$. As a result,
\begin{align}
\label{nr}
&\text{Pr}(\tau<\rho)\tl{\tau}_\text{min}(s) = \langle e^{-s\tau} \rangle,\nonumber\\
&= \int^\infty_0 dt~f_\tau(t) \int^\infty_t dt'~f_\rho(t') e^{-st},\nonumber\\
&= \Big(\int^T_0 + \int^\infty_T\Big)dt~f_\tau(t)e^{-st}\int^\infty_t dt'~\delta(t'-T),\nonumber\\
&= \int^T_0 dt~f_\tau(t)e^{-st}.
\end{align}
The $\int^\infty_T$ integral in the third line does not contribute anything as the
limits of integration do not contain the point $t' = T$. In a similar manner
\begin{align}
\label{dr}
\text{Pr}(\rho\le\tau)\tl{\rho}_\text{min}(s) = e^{-sT}\int^\infty_\tau dt~f_T(t).
\end{align}
Using (\ref{nr}) and (\ref{dr}) in Eq.~(\ref{F_res}) we find that the FPTD of a
stochastic process under sharp restart is given by
\begin{align}
\label{fptd_sh}
\tl{F}_T(s) = \fr{\int^T_0 dt~f_\tau(t)e^{-st}}{1-e^{-sT}\int^\infty_T dt~f_\tau(t)},
\end{align}
where $f_\tau(t)$ is the FPTD without restart and the subscript $T$ on the lhs denotes
the time of sharp restart $T$. From the FPTD in (\ref{fptd_sh}) follows the MFPT under
sharp restart
\begin{align}
\label{mfpt_s}
\avgt = -\fr{d}{ds}\tl{F}_T(s)\Big|_{s=0} = \fr{\int^T_0 dt~q_\tau(t)}{\int^T_0 dt~f_\tau(t)},
\end{align}
which has been earlier derived in Ref.~\cite{eliazar2020mean} in an alternative
manner with $q_\tau(t)$ denoting the survival probability.

For the system of two vicious random walkers $f_\tau(t) = F(t)
= \fr{L}{\sqrt{4\pi D_r t^3}}\exp\Big(-\fr{L^2}{4D_r t}\Big) \Rightarrow q_\tau(t) = \text{erf}\Big(
\fr{L}{\sqrt{4D_r t}}\Big)$ \cite{redner2001guide}. Using these in (\ref{mfpt_s}) we have
the MFPT to annihilation under sharp resetting
\begin{align}
\avgt = \fr{\sqrt{\fr{L^2 T}{\pi D_r}}e^{-\fr{L^2}{4D_r T}} - \fr{L^2}{2D_r}
\text{erfc}\Big(\fr{L}{\sqrt{4D_r T}}\Big)+T\text{erf}\Big(\fr{L^2}{\sqrt{4D_r T}}\Big)}
{\text{erfc}\Big(\fr{L}{\sqrt{4D_r T}}\Big)}.
\end{align}
The integral of $q_\tau(t)$ above has been evaluated using the integral representation of
the error function and related Laplace transforms \cite{oberhettinger2012tables,gradshteyn2014table}.
Once again it is evident that resetting gives a finite MFPT. Next we look at the PDF
of the intersection point.

\begin{figure}
\includegraphics[width=0.5\textwidth]{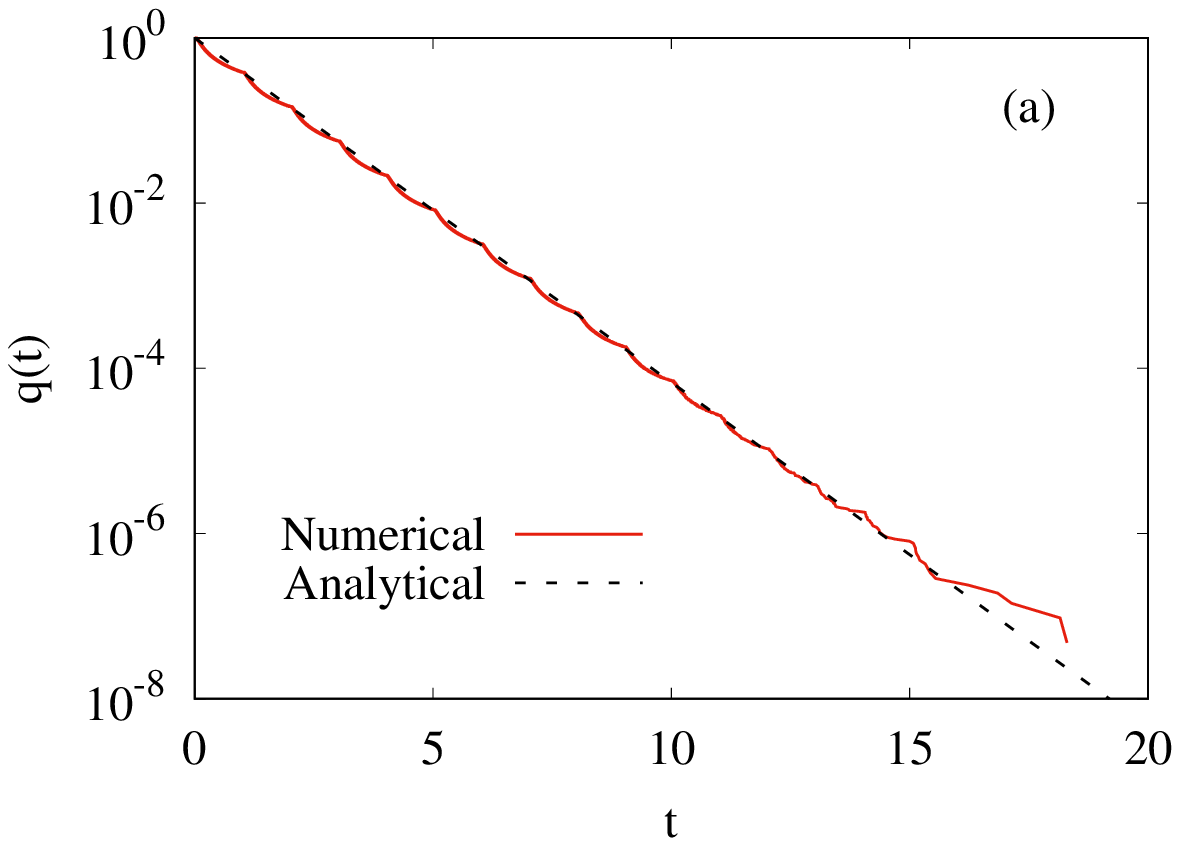}\\
\includegraphics[width=0.5\textwidth]{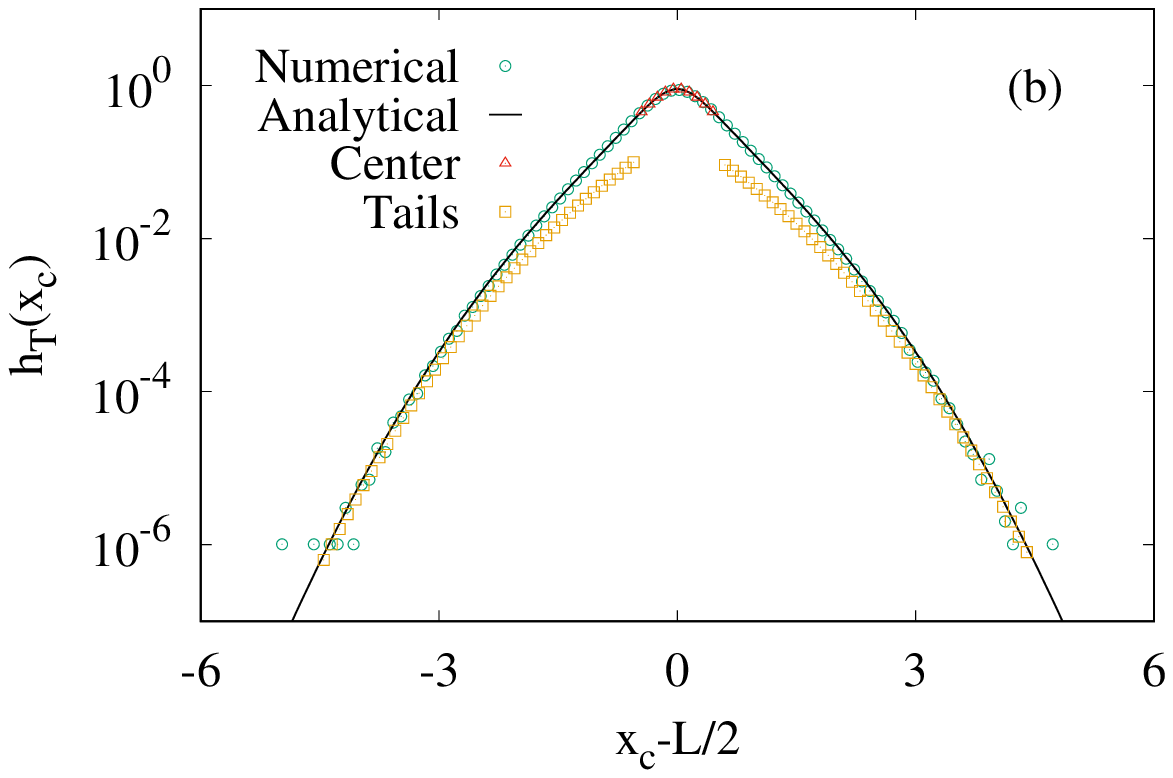}
\caption{(a) Survival probability $q(t)$ for the two vicious walkers
following Eq.~(\ref{dyn}) to not cross paths upto time $t$ sharply reset to their
initial positions after time $T$. Red solid line represents
numerically estimated $q(t)$ while the black dashed line represents the
analytical form: $q(t) \sim \exp(-|s_{0,T}|t)$. (b) Numerically estimated PDF of
the location of intersection $h_T(x_c)$ (green circles) is compared against the
approximate form in Eq.~(\ref{sharp}) (black solid line). Yellow squares denote the contribution
to $h^t_R$ for $|x_c-L/2|>L/2$ and red triangles $h^c_R$ for $|x_c-L/2|<L/2$.
We have used a factor $b$ such that $h_R \sim b h^c_R$ to demonstrate that
$h^c_R$ indeed captures the center of $h_R$ (upto a scale).
Parameter values are: $D_1, ~D_2 = 1,~L = 1,~T = 1$ and $b = 0.4$.}
\label{sharp_fig}
\end{figure}

In order to evaluate the PDF of the intersection point under sharp resetting, we need
the time domain representation of FPTD in Eq.~(\ref{fptd_sh}). Using
$F(t) = \fr{L}{\sqrt{4\pi D_r t^3}}\exp\Big(-\fr{L^2}{4D_r t}\Big)$ we have
\begin{align}
F_T(t) &= \fr{1}{2\pi i}\int^{\gamma+i\infty}_{\gamma-i\infty}ds~
\Big[\fr{\int^T_0 dt~F(t)e^{-st}}{1-e^{-sT}\int^\infty_T dt~F(t)}\Big]~e^{st},\nonumber\\
&\stackrel{\text{large}~t}{\approx} \fr{1}{2\pi i}\int^{\gamma+i\infty}_{\gamma-i\infty}ds~
\Big[\fr{\int^T_0 dt~F(t)}{1-e^{-sT}\int^\infty_T dt~F(t)}\Big]~e^{st},\nonumber\\
&= \fr{1}{2\pi i}\int^{\gamma+i\infty}_{\gamma-i\infty}ds~
\fr{\text{erfc}\Big(\fr{L}{\sqrt{4D_r T}}\Big)}{1-e^{-sT}\text{erf}\Big(\fr{L}{\sqrt{4D_r T}}
\Big)}e^{st}
\end{align}
where we have used the approximation $\int^T_0 dt~F(t)e^{-st} \approx \int^T_0 dt~F(t)$
in the limit of large times $t$ (small $s$ behavior). The integral is now straightforwardly
evaluated from the
pole in the complex plane located at $s_{0,T} = \fr{1}{T}\log \text{erf}\Big(\fr{L}{\sqrt{4D_r T}}
\Big)$. As a result
\begin{align}
F_T(t) \stackrel{\text{large}~t}{\approx} \fr{1}{T}\fr{\text{erfc}\Big(\fr{L}{\sqrt{4D_r T}}\Big)}
{\text{erf}\Big(\fr{L}{\sqrt{4D_r T}}\Big)}e^{s_{0,T}(t+T)}
\end{align}
which implies that at large times $F_T(t)$ decays exponentially as $s_{0,T} < 0$, since
$\text{erf}\Big(\fr{L}{\sqrt{4D_r T}}\Big)<1~\forall~T>0$. We numerically estimate
the characteristic decay time in Fig.~\ref{sharp_fig}(a) and find good agreement with
the analytical result.

Now we estimate the PDF of the center of mass under sharp resetting. When the
two vicious walkers are reset to their initial locations regularly after interval $T$,
the number of renewals taking place upto time
equals $\lfloor \fr{t}{T} \rfloor$ ($\lfloor \rfloor$ denotes the
floor function). Furthermore, since the center of mass starts afresh after every
reset, its PDF at time $t$ is given by
\begin{align}
p_T(x_c,t) = \fr{1}{\sqrt{4\pi D_c \Big(t-\lfloor \fr{t}{T} \rfloor T\Big)}}
\exp\Big[-\fr{(x_c-L/2)^2}{4 D_c \Big(t-\lfloor \fr{t}{T} \rfloor T\Big)}\Big].
\end{align}
For $t < T$ we have $\lfloor \fr{t}{T} \rfloor = 0$ and the center of mass evolves
with the PDF $p(x_c,t)$. If the trajectories of the two walkers cross before
any restart, then the time of their annihilation follows the FPTD $F(t)$. As a
result, similar to the case of Poissonian resetting, the PDF of the
intersection point is composed of two parts: one coming from the trajectories
which annihilate each other at $t < T$, and the remaining ones which undergo
at least one reset event before crossing their paths. Thus, the PDF of the
intersection point under sharp resetting is
\begin{align}
\label{sharp}
h_T(x_c) &= \int^T_0 dt~F(t) p(x_c,t) + \int^\infty_T dt~F_T(t)p_T(x_c,t)\nonumber\\
&\equiv h^c_T(x_c) + h^t_T(x_c),
\end{align}
where $h^c_T$ and $h^t_T$ denote the integrals in the intervals $[0,T]$
and $[T,\infty)$ respectively and are defined analogous to their Poissonian
counterparts. The first integral is relatively straightforward and evaluates to
\begin{align}
\label{sharp_first}
&h^c_T(x_c)\nonumber\\
&= \fr{1}{\pi}\fr{L/\sqrt{D_rD_c}}{\fr{L^2}{4D_r} + \fr{(x_c-L/2)^2}{4D_c}}
\exp\Big[-\fr{1}{T}\Big\{\fr{L^2}{4D_r} + \fr{(x_c-L/2)^2}{4D_c}\Big\}\Big].
\end{align}
This implies that under sharp resetting the fat tails of the Cauchy distribution
are tamed to an effective Gaussian. It becomes even more interesting
once we realize that here we are considering those trajectories which
have not even reset once. In other words, the fact that a restart is set to
take place at $t = T$, forces a certain fraction of trajectories to cross their
paths, thus introducing Gaussian cutoffs in the tails. In addition,
the central part of the PDF close to the initial location of
the center of mass is Gaussian. Now coming to the second
integral in (\ref{sharp}), we have
\begin{align}
\label{sharp_sec}
&h^t_T(x_c)\nonumber\\
&\approx \fr{1}{\sqrt{4\pi D_c T^2}} \fr{\text{erfc}\Big(\fr{L}
{\sqrt{4D_r T}}\Big)}{\text{erf}\Big(\fr{L}{\sqrt{4D_r T}}\Big)} e^{-|s_{0,T}|T}\nonumber\\
&\times \int^\infty_T dt~\fr{\exp\Big[-\fr{(x_c-L/2)^2}{4 D_c \Big(t-\lfloor
\fr{t}{T} \rfloor T\Big)} -|s_{0,T}|t\Big]}{\sqrt{t-\lfloor \fr{t}{T} \rfloor T}},
\end{align}
where the limits of integration are kept from $T$ to $\infty$ for reasons stated
above. The integral in (\ref{sharp_sec}) can be evaluated by decomposing the interval
of integration into subintervals of length $T$. This helps us to reduce the integral
on the real line $[T,\infty)$ to an integration over the interval $[0,T]$. The
reason we can do this is that the floor function turns $t-\lfloor \fr{t}{T}
\rfloor T$ into a periodic function. As a result
\begin{align}
\label{intl}
&\int^\infty_T dt~\fr{\exp\Big[-\fr{(x_c-L/2)^2}{4 D_c \Big(t-\lfloor
\fr{t}{T} \rfloor T\Big)} -|s_{0,T}|t\Big]}{\sqrt{t-\lfloor \fr{t}{T} \rfloor T}}\nonumber\\
&= \sum^\infty_{m=1} e^{-|s_{0,T}|mT}\int^T_0 dw \fr{1}{\sqrt{w}}e^{-\fr{a}{w}-|s_{0,T}|w}\nonumber\\
&= \fr{\sqrt{\pi/|s_{0,T}|}}{e^{|s_{0,T}|T}-1}\Big[e^{-2\sqrt{a|s_{0,T}|}}
-\fr{e^{-2\sqrt{a|s_{0,T}|}}}{2}\text{erfc}
\Big(\fr{T\sqrt{|s_{0,T}|}-\sqrt{a}}{\sqrt{T}}\Big)\nonumber\\
&-\fr{e^{2\sqrt{a|s_{0,T}|}}}{2}\text{erfc}
\Big(\fr{T\sqrt{|s_{0,T}|}+\sqrt{a}}{\sqrt{T}}\Big)\Big]
\end{align}
where $a = \fr{(x_c-L/2)^2}{4D_c}$ and the integral is evaluated using MAXIMA.
This implies that the PDF of
the intersection point has exponentially decaying tails for large $a$. Using (\ref{intl})
in (\ref{sharp_sec}) and then along with (\ref{sharp_first}) in Eq.~(\ref{sharp})
we have
\begin{widetext}
\begin{align}
h_T(x_c)\approx\begin{cases}
\fr{1}{\pi}\fr{L/\sqrt{D_rD_c}}{\fr{L^2}{4D_r} + \fr{(x_c-L/2)^2}{4D_c}}
\exp\Big[-\fr{1}{T}\Big\{\fr{L^2}{4D_r} + \fr{(x_c-L/2)^2}{4D_c}\Big\}\Big],~\text{center},\\
\fr{\sqrt{\pi/|s_{0,T}|}}{\sqrt{4\pi D_c T^2}} \fr{\text{erfc}\Big(\fr{L}
{\sqrt{4D_r T}}\Big)}{\text{erf}\Big(\fr{L}{\sqrt{4D_r T}}\Big)}
\fr{e^{-|s_{0,T}|T}}{e^{|s_{0,T}|T}-1}\Big[e^{-2\sqrt{a|s_{0,T}|}}
-\fr{e^{-2\sqrt{a|s_{0,T}|}}}{2}\text{erfc}
\Big(\fr{T\sqrt{|s_{0,T}|}-\sqrt{a}}{\sqrt{T}}\Big)
-\fr{e^{2\sqrt{a|s_{0,T}|}}}{2}\text{erfc}
\Big(\fr{T\sqrt{|s_{0,T}|}+\sqrt{a}}{\sqrt{T}}\Big)\Big],~\text{tails}.
\end{cases}
\end{align}
\end{widetext}
We numerically study the PDF of the intersection point in Fig.~\ref{sharp_fig}(b)
and find that Eq.~(\ref{sharp}) agrees well with numerical calculations. Furthermore,
the PDFs $h^c_T$ (with a scale factor $b$) and $h^t_T$ also individually agree
with the numerically estimated $h_T$ in their respective ranges (as stated above
in case of Poissonian resetting). While it
may not be apparent from Fig.~\ref{sharp_fig}
(b), the tails of $h_T$ are indeed exponential. This follows from the fact that in Eq.~(\ref{intl})
the second erfc term describing the tails approaches a constant for large fluctuations,
while the third term approaches zero. As a consequence, $h^t_T \sim e^{-2\sqrt{a|s_{0,T}|}}$
for large $|x_c-L/2|$.

\section{Comparing Poisson resetting and sharp resetting}
\begin{figure}
\includegraphics[width=0.5\textwidth]{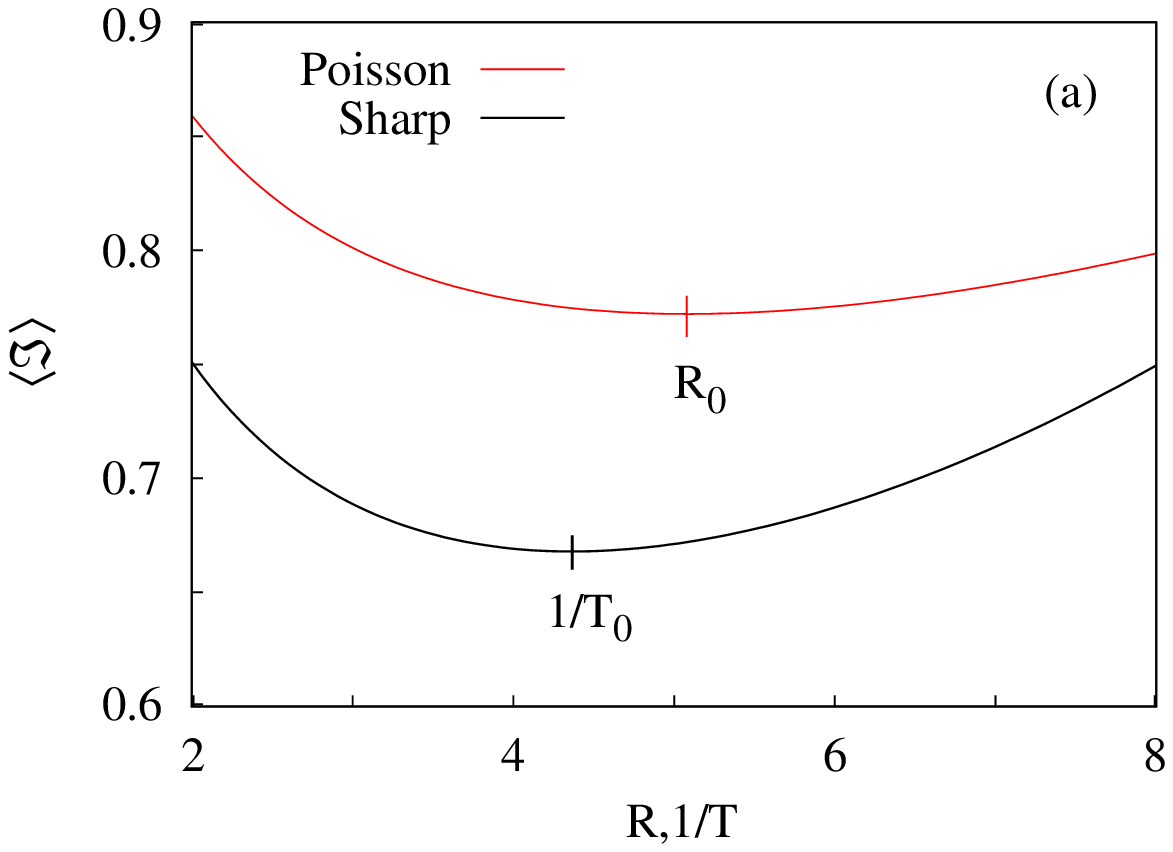}\\
\includegraphics[width=0.5\textwidth]{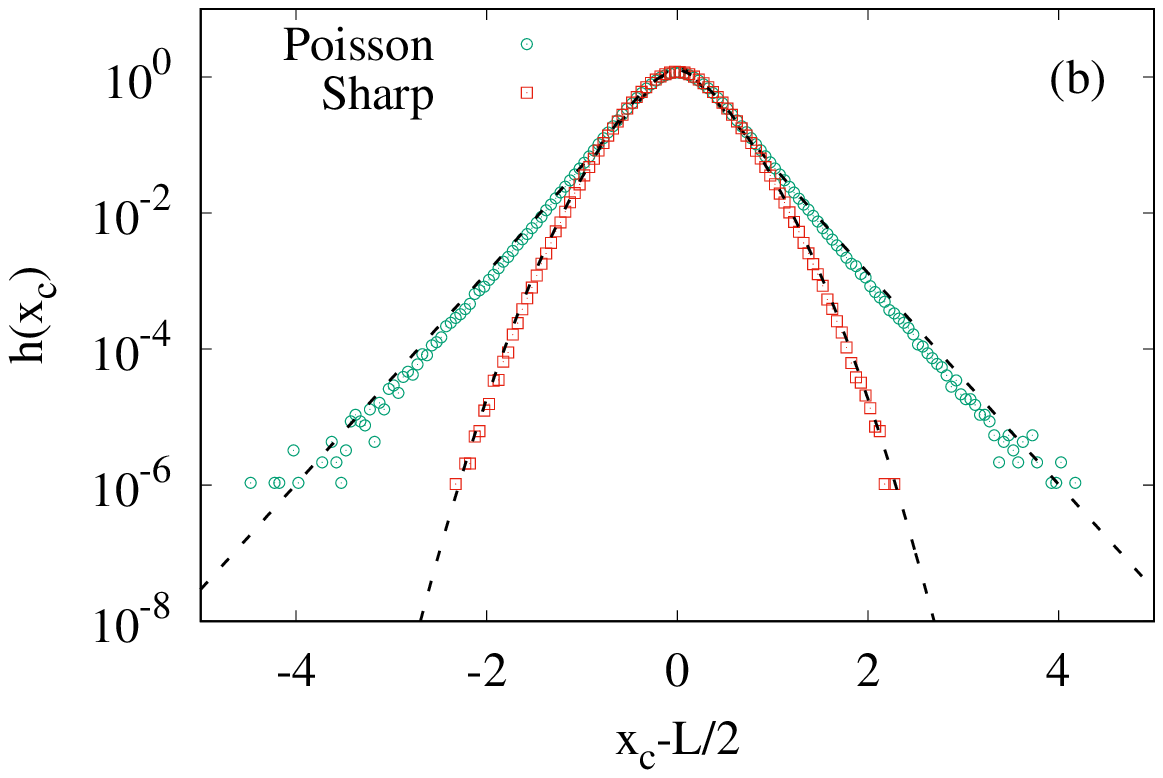}
\caption{(a) MFPT $\langle \mathscr{T} \rangle$ for the system of
two vicious walkers under Poisson and sharp resetting. The location
of the minima are marked indicating the optimal resetting rate $R_0$
and the optimal time of sharp reset $T_0$. (b) The PDF of the location
of annihilation $h(x_c)$ for the two resetting protocols at their
optimal values. Parameter values are:
$D_1, ~D_2 = 1,~L = 1,~R_0 = 5.079$ and $T_0 = 0.229$.}
\label{compare}
\end{figure}
So far we have studied the dynamics of two vicious walkers under Poissonian and
sharp resetting, but in separate scenarios. It thus becomes interesting to comparing
the two protocols against each other. Answer to this question is known partly
in that sharp resetting wins over Poissonian resetting in renewal resetting
scenario \cite{pal2017first}. Hence we compare the minima of the MFPTs under
the two resetting protocols. For this purpose let us choose $D_1, ~D_2 = 1$
and $L = 1$. As a result we have
\begin{subequations}
\begin{align}
\avgr &= \fr{\exp(\sqrt{R/2})-1}{R},\\
\avgt &= \sqrt{\fr{T}{2\pi}}\fr{\exp(-1/8T)}{\text{erfc}(1/\sqrt{8T})}
+T\fr{\text{erf}(1/\sqrt{8T})}
{\text{erfc}(1/\sqrt{8T})} - \fr{1}{4}.
\end{align}
\end{subequations}
From the above equations, the minima of the MFPT $\avgr$ occurs at $R=R_0$ where
$\fr{d}{dR}\avgr|_{R=R_0} = e^{\sqrt{R_0/2}}/\sqrt{8R^3_0} - \langle \mathscr{T}_{R_0}
\rangle/R_0 = 0 \Rightarrow R_0 \approx 5.079$. For this value of the resetting
rate $\langle \mathscr{T}_{R_0} \rangle \approx 0.772$. Similarly, $\avgt$ has a
global minima at $T_0 \approx 0.229$ which implies $\langle \mathscr{T}_{T_0}
\rangle \approx 0.668$.
To get a perspective of these numbers, the diffusive time scale of the relative
coordinate to cover a distance $L$ is $\langle \mathscr{T}_D \rangle = L^2/2D_r
= 0.25$ for $L = 1$. On this time scale, the MFPT for the two vicious walkers to
annihilate each other is $\avgr \approx \avgt \approx 3\langle \mathscr{T}_D
\rangle$. Furthermore, the relative advantage of sharp resetting over Poissonian
resetting is $|\avgt-\avgr|/\avgt \approx 0.16$ which is significant.
In other words, while it is suitable for the lion to quickly hunt that both
the lion and the lamb return to their homes after fixed time intervals, for the
lamb Poissonian resetting is better as it might survive a little longer. We
compare the two resetting protocols graphically in Fig.~\ref{compare}(a) and
see the relative advantage of sharp resetting over Poisson resetting.

Let us now look at the tail behavior of the PDF $h^t$ of the intersection point
for Poisson and sharp resetting at their optimal levels respectively. From
Eq.~(\ref{second_int}) it is clear that for Poissonian resetting $h^t_{R_0} \sim
\exp\Big(-\sqrt{\fr{|s_{0,{R_0}}|+R_0}{D_c}}\Big|x_c-\fr{L}{2}\Big|\Big) \approx
\exp(-1.78|x_c-1/2|)$. On the other hand, for sharp resetting we have
$s_{0,T_0} = \fr{1}{T_0}\log\text{erf}\Big(\fr{L}{\sqrt{4D_r T_0}}\Big) \approx
-1.5332$. As a result, the tail part of the PDF of the location of intersection
is $h^t_{T_0} \sim \exp\Big(-\sqrt{\fr{|s_{0,T_0}|}{D_c}}|x_c-L/2|\Big) \approx
\exp(-1.75|x_c-1/2|)$. This implies that at optimal resetting, the tails of
the PDF decay faster for sharp resetting as compared to Poissonian resetting.
This also follows from the fact that at optimal resetting
$\langle \mathscr{T}_{T_0} \rangle < \langle \mathscr{T}_{R_0} \rangle$, as a
result both the lion and the lamb do not venture far from their homes at
the time of capture under sharp resetting as compared to Poissonian resetting.
We compare the two PDFs both numerically and analytically in Fig.~\ref{compare}
(b) and find the PDF for Poissonian resetting has a higher spread as compared
to that for sharp resetting.

\section{Conclusions}
In the realm of nonequilibrium statistical physics vicious random walkers are
used to model interfacial wetting in $1+1$ dimensions and non-intersecting polymers.
In these contexts the survival probability and the distribution of the location
of coalescence are relevant quantities to address. Within the domain of capture problems,
vicious random walks translate to the capture of a prey by a predator. Motivated by
these examples, in this paper we study the annihilation properties of two vicious
random walkers under Poissonian and sharp resetting protocols. In absence of resetting
the mean time of capture is divergent while the location of annihilation follows
a Cauchy distribution. Introduction of resetting in the system renders finite MFPT
due to the fact that the FPTD tails now decay exponentially as compared to algebraically
in absence of resetting. Furthermore, tails of the PDF of annihilation location
now decay exponentially. This
is independent of the exact nature of resetting protocol. The central part of the
PDF, however, depends on the way system is reset to its initial location. For
Poissonian resetting the central part of the PDF is a Cauchy distribution, while
for sharp resetting it is a Gaussian.

We have reset the two walkers identically so that we can reduce the two particle
system as to be described by the motion of the center of mass and motion about the
center of mass. We have also assumed that restarts are instantaneous, but in any
realistic scenario bringing back the system to its initial state takes a finite amount
of time. Even within the realm of instantaneous resetting, we chose
the particles to be identical. What would happen if we include inertia and assign
different masses to different particles? We explore these and other interesting
possibilities in future works.

\textit{Acknowledgments}: RKS thanks the Israel Academy of Sciences and Humanities (IASH)
and the Council of Higher Education (CHE) Fellowship. SS acknowledges the HPC facility at Ben-Gurion 
University.  

\bibliographystyle{apsrev4-1}
\bibliography{ref.bib}

\end{document}